\documentclass[a4paper, 12pt]{article}

\usepackage{graphicx}
\usepackage{url}
\usepackage{amsbsy,amssymb} 
\usepackage{amsmath}
\usepackage{abstract}
\usepackage{hyperref}
\usepackage{color}

\usepackage{amsmath,amsthm}

\newtheorem{Proposition}{Proposition}

\newtheorem{Theorem}{Theorem}
\newtheorem*{Proof}{Proof}

\newtheorem{Definition}{Definition}

\newtheorem{Assumption}{Assumption}
\newtheorem{Axiom}{Axiom}

\begin{document}

{\LARGE\centering{\bf{Entropy in Thermodynamics: from Foliation to Categorization}}}

\begin{center}
\sf{Rados\l aw A. Kycia$^{1,2,a}$}
\end{center}

\medskip
\small{
\centerline{$^{1}$Masaryk University}
\centerline{Department of Mathematics and Statistics}
\centerline{Kotl\'{a}\v{r}sk\'{a} 267/2, 611 37 Brno, The Czech Republic}
\centerline{\\}
\centerline{$^{2}$Cracow University of Technology}
\centerline{Faculty of Materials Engineering and Physics}
\centerline{Warszawska 24, Krak\'ow, 31-155, Poland}
\centerline{\\}

\centerline{$^{a}${\tt
kycia.radoslaw@gmail.com}}
}

\begin{abstract}
\noindent
We overview the notion of entropy in thermodynamics. We start from the smooth case using differential forms on the manifold, which is the natural language for thermodynamics. Then the axiomatic definition of entropy as ordering on a set that is induced by adiabatic processes will be outlined. Finally, the viewpoint of category theory is provided, which reinterprets the ordering structure as a category of pre-ordered sets.
\end{abstract}

\begin{center}
For Professor Olga Rossi, in memoriam.
\end{center}

\section{Introduction}
\begin{figure}
\centering
 \includegraphics[width=0.5\textwidth]{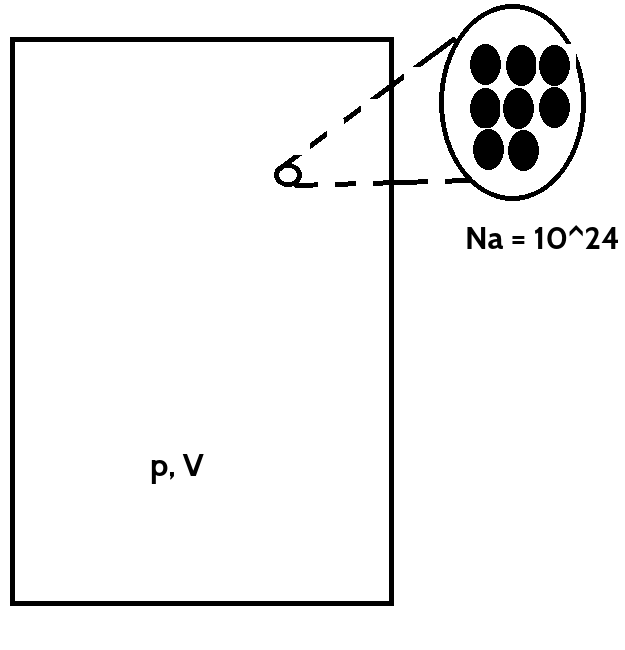}
 \label{Fig.DimensionReduction}
 \caption{System of large degrees of freedom (atoms) in thermodynamics is reduced to a few variables.}
\end{figure}

The notion of entropy ('tropos' is Greek word transformation) initially appeared in Thermodynamics to describe the possible direction of the process. At the time the theory was being developed,  conception regarding the inner structure of matter, like atoms, was not available, and hence matter was described in terms of macroscopic averaged variables as pressure, volume, temperature, etc. Currently, we know that these variables come from the reduction of a large number of degrees of freedom of particles in a piece of matter (the Avogadro constant $N_{A}\sim 10^{23}$ atoms which in classical description have $3$ numbers describing position coordinates, and $3$ numbers describing velocity coordinates) to a few variables mentioned above. The need for pointing out this 'coarse/average' evolution direction was imminent, and to fulfill this need entropy was invented, see Fig. \ref{Fig.DimensionReduction}. 
Does it mean that if an entropy in theory appears, then we are dealing with a 'coarse' (not fundamental) description? We do not know yet. 

The history of thermodynamics is full of amazing reasonings that finally lead to the correct laws of nature. To mention one, Robert Mayer concluded that heat is energy flow by observing the color of the blood of sailors under different geographic latitudes on the ship on which he was a medical doctor.

Since the pioneering work of Boltzmann that connects thermodynamical entropy with microscopic properties of matter and properties of logarithm function, this notion appeared to be useful in various disciplines like information theory \cite{InformationEntropy} or dynamical systems \cite{Katok}. The increasing importance of this concept is reflected in bibliometric data of research papers on this subject \cite{Entropy_BibliometricAnalysis}.

This summary will neglect all the classical/physical motivation for thermodynamics, and we go directly to mathematical concepts associated with the notion of entropy. We believe that thanks to such an approach, we avoid mixing assumptions, the result of reasoning, and 'common knowledge' in this theory, which is common in physics literature and which leads to the difficulty in grasping these concepts. Thermodynamics is mature enough to axiomatize fully, and this can be done as will be presented below. This presentation is by no means original research - it is only an overview of the subject and a small part of existing literature. Only the organization of the material is perhaps nonstandard and selective. However, it is hoped it can be treated as a guide for novices (both with mathematical or physical background) to avoid pitfalls common in this subject.

The overview is organized as follows: First, the geometric meaning of entropy close to the original formulation in modern differential geometric terms will be provided. The presentation will be provided with the context, i.e., geometric structure of (phenomenological/equilibrium) thermodynamics. Then the axiomatic approach to entropy will be outlined. Finally, the categorical approach to this subject will be presented. In the Appendix, mathematical preliminaries were collected for the reader's convenience, and we advise the reader to look up the Appendix to be oriented what kind of mathematics is needed to understand the main parts of the paper.

The Ostrava Seminar on Mathematical Physics organized for many years by Diana Barseghyan, Olga Rossi, and Pasha Zusmanovich is a unique platform to exchange knowledge between mathematicians and physicists. There is also a big audience of students that motivates speakers to present material more pedagogically, including also the context of the research. One of my talks, which was a pleasure to deliver, was about entropy and Landauer's principle. This overview paper can be treated as a basic introduction and guide to the subject.

\section{Smooth case}
We start by describing thermodynamics in the natural setup of smooth manifolds with contact structure. In these terms, although not so precisely due to less developed mathematical language, the fathers of thermodynamics were thinking. The presentation follows closely \cite{LychagingThermodynamics, GeometryOfPhysicsFrankel, CallenThermodynamics, BabergSternberg, BoylingThermodynamics, IngardenJamiolkowski}.

\subsection{Space}
In thermodynamics, we identify some system from the environment by distinguishing some more or less formal boundaries with some specific physical properties (e.g., heat contact or permeability of particles). Such a system should be macroscopically uniform in the sense of its physical and chemical properties - a so-called \emph{simple system}. By distinguishing such a system, we can describe it by some variables depending on the physical context. The common feature of these variables is their uniform behavior under the scaling group $\mathbb{R}^{+}$, which reflects the physical property that the scaling of the system scales its internal parameters, e.g., scaling the system scales its volume or energy. These variables are called \emph{extensive}. Call these variables $\{X_{i}\}_{i=0}^{n}$. If there is another extensive variable $X$, then it must depend on the previous $X=X(X_{0},\ldots, X_{n})$ and scales as $X(\lambda X_{0},\ldots, \lambda X_{n})=\lambda X(X_{0},\ldots, X_{n})$ where $\lambda \in \mathbb{R}^{+}$ that it should be truly extensive.

The common choice of extensive variables and the usual symbols (instead of $X_{i}$'s) designated for them are as follows:
\begin{itemize}
 \item {$U$ - energy of the system;}
 \item {$V$ - volume;}
 \item {$N$ - number of particles;}
\end{itemize}

The first assumption is that 
\begin{Assumption}
 In an equilibrium state, the system is fully described by some set of extensive variables.
\end{Assumption}
An equilibrium state is attained when the system is left on its own and relaxes attaining this state without any further change of extensive variables. The direction in which a system left on its own evolves is described by entropy, which will be introduced at a later stage.

If a system is composed of more simple systems, then the number of variables multiplies accordingly.

The system (simple or compound) can interact with the environment by exchanging energy. One 'directed' way of transferring energy is work made by the environment on the system. Various types of work are described by work 1-forms that relate change of extensive parameters of the system with work done on the system. In local coordinates:
\begin{equation}
 W = P_{i}dX^{i}.
\end{equation}
The coefficients $P_{i}$ are called \emph{intensive} variables and describe the 'generalized forces' of environment that act on the system. They do not scale. Note that  $W$ does not to be an exact form, and therefore work may depend on the path along which it is integrated.

Common choice of intensive variables are
\begin{itemize}
 \item {$p$ - pressure; associated with change of volume $V$;}
 \item {$-\mu$ - chemical potential describing density of work done by changing the number of particles in the system by adding/removing particles/elements from/to environment or modified by chemical reactions; associated with the number of particles $N$;}
\end{itemize}

The second way of energy transfer between system and environment is the heat transfer described by a 1-form $Q$.  We will see below that $Q$ can be written in terms of work form: $Q=TdS$, where $T$ is the absolute temperature (intensive variable), and $S$ is the entropy (extensive variable). In physical terms, $Q$ is the transfer of kinetic energy at the level of atoms/molecules. The detailed description of this transfer in thermodynamics is neglected by reducing microscopic degrees of freedom to a few macroscopic ones. Therefore some additional law has to be introduced that controls such transfer. This is done in terms of entropy and the Second Law of Thermodynamics.

In thermodynamics, the system is described by energy $U$ and $2n$ pairs of associated intensive-extensive variables. These are local coordinates on $2n+1$ dimensional manifold. From physics, it is assumed
\begin{Assumption}
 The equilibrium state is described as a point in $2n+1$ dimensional smooth manifold $M$ called the space of states.
\end{Assumption}
Local coordinates are usually taken to be $(U, (T, S), (p,V), (\mu, N), \ldots)$, where intensive-extensive pairs were grouped.

In order to compare systems in equilibrium, we introduce the Zeroth Law of Thermodynamics
\begin{Axiom}
 If a system $A$ is with a thermal equilibrium with $B$ and $B$ with $C$, then $A$ is in thermal equilibrium with $C$.
\end{Axiom}
The thermal equilibrium of two systems means that there is no heat flow $Q$ between systems that are connected by thermally conducting material. The Law means that the relation of 'being in thermal equilibrium' is transitive. It is also obviously reflexive and symmetric, and therefore is an equivalence relation. It allows us to define tools/systems called thermometers that measure \emph{empirical temperature}, which represent precisely these equivalence classes. This empirical temperature will be related to the (absolute) temperature $T$ below.

\subsection{Processes}
The next step is to consider the change in the system that is described by paths in the space of states called thermodynamical processes.

(Equilibrium) thermodynamics is only occupied with \emph{quasi-static processes}, which can be represented by curves in the space of states. In physical terms, they can be considered as physical/chemical processes that occur 'slow enough' that in every step of the process the system and environment are in equilibrium or relax 'fast enough' to equilibrium. It is only an idealization. On the other hand, \emph{non-quasi-static processes} cannot be described as a path in the space of states. They can only be marked as initial and final points if these points are equilibrium states. This peculiarity is connected with the fact that the points in the space of states describe only equilibrium states.

The other distinction is according to reversibility. The process is:
\begin{itemize}
 \item {\emph{reversible} - if it can be conducted in both directions when all variables (intensive of system and extensive of environment) can be returned to initial values in local description;}
 \item  {\emph{irreversible} - it cannot be reversed;}
\end{itemize}

For quasi-static processes we have a curve $\gamma$ in the space of states $M$ that we assume to be piecewise smooth which is usual assumption. We can then calculate:
\begin{itemize}
 \item {$\Delta Q (\gamma) := \int_{\gamma} Q$ - total heat transfer in the process;}
 \item {$\Delta W (\gamma) := \int_{\gamma} W$ - total work done in the process;}
\end{itemize}
Note that these definitions are not valid when there is no curve along which 1-forms $Q$ and $W$ can be integrated, i.e. for non-quasi-static processes.

Some examples of thermodynamic processes are as follows \cite{GeometryOfPhysicsFrankel}:
\begin{itemize}
 \item {\emph{Quasi-static adiabatic process} - in this case no heat is exchanged, that is $\Delta Q(\gamma)=0$;}
 \item {\emph{Heating at constant volume} - a quasi-static process which for the case of simple system takes place without exchange of particles $\Delta W (\gamma) = \int_{\gamma} pdV=0$;}
 \item {\emph{Non-quasi-static process} - no path in $M$ therefore no $\Delta W$ and no $\Delta W$ can be calculated. Only the the difference of energy between initial and final state of the process can be defined.}
\end{itemize}

In technical applications, the most important are closed paths that are called \emph{thermodynamical cycles} and describe the cyclic work of engines. They are also crucial in the formulation of the Second Law of Thermodynamics below.

\subsection{The First Law of Thermodynamics}
The first fundamental law of thermodynamics describes from the physical point of view the conservation of energy during a quasi-static process, namely
\begin{Axiom} (First Law of Thermodynamics)
 \begin{equation}
  dU = Q-W
  \label{Eq.FirstLawOfThermodynamics}
 \end{equation}
\end{Axiom}
We stated this law as an Axiom since, although on the physics side it is a fundamental law of nature,  on the mathematical side it is an unquestionable statement, i.e., an axiom for mathematical formulation of thermodynamics.

For quasi-static processes described by the curve $\gamma$ in $M$  with the initial point $x$ and the final point $y$ the integrated version of (\ref{Eq.FirstLawOfThermodynamics}) is
\begin{equation}
 \Delta U (\gamma) := U(y)-U(x)=\Delta Q(\gamma)-\Delta W(\gamma).
\end{equation}
This is due to the fact that $dU$ is exact form and therefore its integral depends only on the endpoints of the curve $\gamma$.

In expanded form (\ref{Eq.FirstLawOfThermodynamics}) can be written in local coordinates as
\begin{equation}
 dU = Q -pdV-\mu dN.
\end{equation}

In this context we can reinterpret the properties of the processes:
\begin{itemize}
 \item {\emph{Quasi-static adiabatic process} - $\Delta Q(\gamma)=0$ and therefore $\Delta U = -\Delta W$;}
 \item {\emph{Heating at constant volume} - $\Delta W (\gamma) = \int_{\gamma} pdV=0$ and therefore $\Delta U = \Delta Q$;}
\end{itemize}
Note that a quasi-static adiabatic process converts all the total energy of the system to the work that can be extracted from or transferred to the system. The restrictions on this process prevents the construction of a 'perpetuum mobile' and is controlled by the Second Law of Thermodynamics described below.

We now turn to finishing the mathematical description of state space. On $2n+1$ dimensional space $M$ we have the form
\begin{equation}
 \theta:=dU-Q+W.
 \label{Eq.FirstLawOfThermodynamics_ContactForm}
\end{equation}
The volume form in $M$ can be given by 
\begin{equation}
 \theta \wedge (d \theta)^{n} \neq 0.
\end{equation}
Therefore $\theta$ defines a contact structure on $M$ or equivalently $J^{1}(N)$, where $N$ (see Appendix) has local coordinates as extensive variables $(U,V,N)$. This leads to the final definition of the space of states for thermodynamics
\begin{Definition}
 The space of states in thermodynamics is described by odd dimensional space $M$ with contact form $\theta$ that fulfills $\theta \wedge (d \theta)^{n} \neq 0$.
\end{Definition}
We can now reconstruct the conservation law of the First Law of Thermodynamics: Using the Darboux theorem for contact manifolds (see Appendix), there are local coordinates $(X_{0}, (X_{1},P_{1}), \ldots, (X_{n},P_{n}))$ that the canonical form of $\theta$ is
\begin{equation}
 \theta = dX_{0}-\sum_{i=1}^{n}P_{i}dX^{i}.
 \label{Eq.ContactForm_Canonical_Thermodynamics}
\end{equation}
Comparing with (\ref{Eq.FirstLawOfThermodynamics_ContactForm}) we have that $X_{0}=U$ etc.

In this space the submanifold $\Phi$ describing the physical system in equilibrium fulfills
\begin{equation}
 \Phi^{*}\theta=0,
\end{equation}
that is, physical systems are described by such submanifolds of $M$ that preserve energy/The First Law of Thermodynamics. In the case of a non-degenerate thermodynamical system, it is assumed:
\begin{Assumption}
 The non-degenerate thermodynamical system is described by maximal dimension subspace in the contact space of states of dimension $2n+1$, i.e., Legendre submanifolds of dimension $n$.
\end{Assumption}

The Legendre submanifold is defined by providing $X_{0}=X_{0}(X_{1},\ldots,X_{n})$. Alternatively, using (\ref{Eq.ContactForm_Canonical_Thermodynamics}), we can provide \emph{equations of state}
\begin{equation}
 \left\{ 
 \begin{array}{l}
  P_{1}=P_{1}(X_{1},\ldots,X_{n})=\frac{\partial X_{0}}{\partial X_{1}} \\
  \ldots \\
  P_{n}=P_{n}(X_{1},\ldots,X_{n})=\frac{\partial X_{0}}{\partial X_{n}}.
 \end{array}
\right.
\end{equation}
This can be viewed as the equivalence of holonomic sections of jet space and Legendre submanifolds on contact space - see Appendix.

The last remaining issue is the direction of heat transfer, which is resolved by the Second Law of Thermodynamics outlined in the next subsection.

\subsection{The Second Law of Thermodynamics}
We now put some restrictions on quasi-static adiabatic paths/processes $\gamma$ that are described by $\gamma^{*}Q=0$. All tangent vectors to such paths are in $Ker(Q)$ and define some distribution on $M$. Since adiabatic processes along arbitrary paths are not present in nature, therefore Caratheodory formulated the following version of the Second Law of Thermodynamics:
\begin{Axiom}\cite{GeometryOfPhysicsFrankel} (Second Law of Thermodynamics, Caratheodory) \\
 In a neighborhood of any state $x\in M$ there is state $y$ that is not accessible from $x$ via quasi-static adiabatic paths $\gamma$ such that $\gamma^{*}Q=0$.  
\end{Axiom}
Using the Caratheodory's theorem on accessibility (see Appendix), we get that the distribution $Ker(Q)$ is integrable (defines holonomic constraints in $M$) or, put another way, 
\begin{equation}
 Q\wedge dQ =0.
\end{equation}
This is also equivalent to the statement that
\begin{equation}
 Q = TdS,
\end{equation}
where $T$ is an integrating factor (a nonsingular function on $M$) called the \emph{absolute temperature}, and $S$ is called the entropy. It means that $S=const$ defines a local leaf of the distribution on which quasi-static adiabatic paths lie.

Consider two simple systems with thermal contact (no adiabatic border). It can be shown that these are in equilibrium if their absolute temperatures $T$ are equal \cite{BabergSternberg}. This defines equivalence classes as in the Zeroth Law of Thermodynamics, and therefore absolute temperature can be used as empirical temperature.

There is a stronger version of this law by Kelvin that implies \cite{GeometryOfPhysicsFrankel} Caratheodory's version, namely,
\begin{Axiom}
\label{Axiom_SecondLawOfThermodynamics} (Second Law of Thermodynamics, Kelvin)\\
In quasi-static cyclic process a quantity of heat cannot be converted entirely into its mechanical equivalent of work.
\end{Axiom}
This version will be used hereafter.

It can be shown \cite{GeometryOfPhysicsFrankel} that the foliation exists globally and is not pathological. It relies on the following
\begin{Proposition}\cite{GeometryOfPhysicsFrankel}
The state $y$ obtained from $x$ by cooling at constant volume ($W=0$) cannot be connected again with $x$ by a quasi-static adiabatic process/path.
\end{Proposition}
\begin{Proof}
As in \cite{GeometryOfPhysicsFrankel}, consider, on the contrary, the two paths from $x$ to $y$ shown in Fig. \ref{Fig.AdiabaticVsCooling}.
\begin{figure}
\centering
 \includegraphics[width=0.3\textwidth]{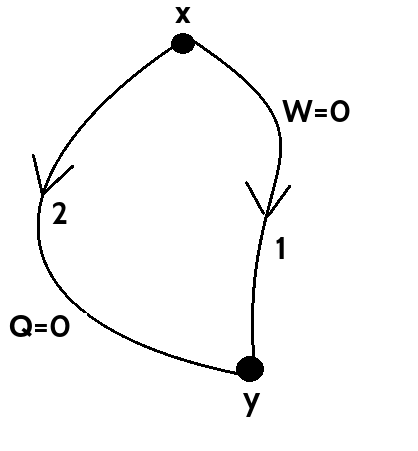}
 \caption{Path $1$ is a cooling at constant volume along which no work is made ($W=0$). The path $2$ is a quasi-static adiabatic process ($Q=0$).}
 \label{Fig.AdiabaticVsCooling}
\end{figure}
The work along $2$ is 
\begin{equation}
 \int_{2}W=\int_{2}Q-dU=-\int_{2}dU= -\int_{1}dU = \int_{-1}dU=\int_{-1}Q,
\end{equation}
as integral of $dU$ does not depend on the path chosen. Here $-1$ means the path is followed in the opposite direction than indicated in the figure. Therefore we have that there is a cycle $y\xrightarrow[-1]{} x \xrightarrow[2]{} y$ which converts whole heat into work, and this contradicts Axiom \ref{Axiom_SecondLawOfThermodynamics}.

Note also that the above proof is, by contraposition,  precisely the statement, that the Second Law of Thermodynamics by Kelvin implies the version of this law by Caratheodory.
\end{Proof}
It results from the above Proposition, that the leaf $S=const$, containing adiabatic processes, is transversal to the paths of the process of cooling at constant volume.  This means that the point starting at some leaf of constant entropy has to be taken into another leaf by the cooling at constant volume. Moving along this path we  never return to the same leaf. This eliminates pathological situations when, e.g., the leaf winds densely on the manifold, i.e., the cases when leaf makes an initial submanifold  \cite{NaturalOperationsInDifferentialGeometry}. This shows that entropy $S$, and $T$ are globally defined on $M$.

The important conclusion that will be a link between classical entropy and its axiomatic definition in the next section is 
\begin{Theorem} \label{Th.SmoothMonotinicity} \cite{GeometryOfPhysicsFrankel}
If a state $y$ results from $x$ by any adiabatic process (quasi-static or not), then $S(y) \geq S(x)$.
\end{Theorem}
We therefore have that in an isolated (i.e., adiabatic) system entropy cannot decrease when achieving equilibrium, which is the commonly known version of the Second Law of Thermodynamics. Here it is presented as a conclusion from a more geometric formulation of this law.

\subsection{Symmetries and thermodynamic potentials}
Having defined the fundamentals of thermodynamics, we can provide some examples of different choices of variables that do not change thermodynamics. They are useful if we prefer to use different variables to observe the system. Since the coordinate changes should not alter the contact structure, they are contact symmetries mentioned in the Appendix. 

The most useful is the Legendre transformation that interchange the role in extensive-intensive pair of variables. This transformation also modifies $X_{0}=U$ variable giving a new, so called, \emph{thermodynamic potential}. We present a few examples in case of constant number of particles (system boundaries are not permeable - $\mu=0$) for simplicity:
\begin{itemize}
 \item {The transformation $p \leftrightarrow V$ gives a thermodynamic potential called the Enthalpy $\tilde{X}_{0}=H:=U+pV$ and the contact form $\tilde{\theta}=dH-TdS-Vdp$. It is useful to observe the system on the submanifold $p=const$.}
 \item {The transformation $T \leftrightarrow S$ gives a thermodynamic potential called the Helmholtz potential/Free energy $\tilde{X}_{0}=F:=U-TS$ and the contact form $\tilde{\theta}=dF+SdT+pdV$. It is useful to observe the system on the submanifold $V=const$.}
 \item {The transformation $p \leftrightarrow V$ and $T \leftrightarrow S$ gives a potential called the Gibbs potential $\tilde{X}_{0}=G:=U+pV-TS$ and the contact form $\tilde{\theta}=dG+SdT-Vdp$. It is useful to observe the system on the submanifold $T=const$ and $p=const$.}
\end{itemize}

\subsection{Examples}
The thermodynamic relations result from taking the exterior derivative of the contact form pulled-back to the Legendre manifold that describes the thermodynamical system. As an example consider the standard contact form
\begin{equation}
 \theta=dU-TdS+pdV.
 \label{Eq.1stLawExample}
\end{equation}
The Legendre manifold $\Phi$ is given by equations of state $T=T(S,V)$ and $p=p(S,V)$. Then since $\Phi^{*}\theta=0$ and $\Phi^{*}d\theta=d\Phi^{*}\theta=0$ we get
\begin{equation}
 \Phi^{*}d\theta = \left( \frac{\partial T}{\partial V} + \frac{\partial p}{\partial S} \right) dS\wedge dV=0,
\end{equation}
which gives one of the Maxwell relations
\begin{equation}
 \frac{\partial T}{\partial V} = -\frac{\partial p}{\partial S}.
 \label{Eq.thermodynamicRelations}
\end{equation}
Since $\Phi$ is also given by $U=U(S,V)$ from (\ref{Eq.1stLawExample}) we get
\begin{equation}
 \Phi^{*}\theta = \left( \frac{\partial U}{\partial S} - T \right) dS + \left( \frac{\partial U}{\partial V} + p \right) dV = 0,
\end{equation}
which means that $T = \frac{\partial U}{\partial S}$ and $p=-\frac{\partial U}{\partial V}$. Then (\ref{Eq.thermodynamicRelations}) can be written as
\begin{equation}
 \frac{\partial^{2}U}{\partial S \partial V} = \frac{\partial^{2}U}{\partial V \partial S},
\end{equation}
which is a tautology for smooth $U$. In general the Maxwell relations can be used as a consistency check of equations of motion - if they define a Legendre submanifold.

Another example is the ideal gas which has the equation of state
\begin{equation}
 pV=NRT,
\end{equation}
where $N$ is the number of moles of the gas, and $R$ is the universal gas constant. This is not enough for the definition of a Legendre submanifold, and another relation is provided
\begin{equation}
 U=\frac{3}{2}NRT.
\end{equation}
These equations are provided for the Lagrange manifolds given by $S=S(U,V)$, which gives
\begin{equation}
 \frac{\partial S}{\partial U}=\frac{1}{T}=\frac{3NR}{2U}, \quad \frac{\partial S}{\partial V}=\frac{p}{T}=\frac{NR}{V}.
\end{equation}
One can easily check that the mixed second derivatives agree.

For more examples, one can look e.g., into \cite{LychagingThermodynamics, BabergSternberg} or for more physical view \cite{CallenThermodynamics}.

\section{Axiomatic approach}
The above description of entropy can be axiomatized. Our presentation in this section closely follows \cite{EntropyOrdering} and \cite{EntropyOrdering2}.

We start from the definition of a simple system, as in the previous section. The states of such a system are points $X,Y,Z, \ldots$ inside the space of states $\Gamma$. Then we fix on the set $\Gamma$ the structure of the space $\mathbb{R}^{2n+1}$, where one variable is the energy $U$ and the remaining $2n$ variables are extensive-intensive pairs.

On such a space we can introduce a scaling by $\lambda, \mu \in \mathbb{R}^{+}$ that is a multiplication group action $\Gamma^{1}=\Gamma$, $(\Gamma^{\lambda})^{\mu}=\Gamma^{\lambda\mu}$. The scaled state $\lambda X$ consists of all extensive variables scaled and all intensive variables unaffected. Two systems $\Gamma_{1}$ and $\Gamma_{2}$ can be composed, and then the composed system is described by points from the Cartesian product $\Gamma_{1}\times \Gamma_{2}$.

The fundamental notion needed for the definition of entropy in \cite{EntropyOrdering, EntropyOrdering2} is adiabatic accessibility
\begin{Definition}
 State $Y$ is \emph{adiabatically accessible} from $X$ if the only result of the transition is a work done. We denote it $X \prec Y$.
\end{Definition}
This definition does not involve heat since it was not defined yet. Besides, the relation $\prec$ is intended to be some 'ordering' to be specified later.

We can further define
\begin{Definition} 
 \begin{itemize}
  \item {\emph{Irreversible adiabatic process}: $X \prec\prec Y$ if $X \prec Y$ and not $Y \prec X$;}
  \item {\emph{Adiabatic equivalence}: $X \sim Y$ if $X \prec Y$ and $Y \prec X$;}
 \end{itemize}
\end{Definition}

In order to introduce entropy $S:\Gamma \rightarrow \mathbb{R}$ the relation $\prec$ is assumed to fulfill the axioms \cite{EntropyOrdering, EntropyOrdering2}:
\begin{itemize}
\item {Monotonicity: $X \sim X$ }
 \item {Transitivity: If $X \prec Y$ and $Y \prec Z$ then $X \prec Z$}
 \item {Consistency: $X \prec X'$ and $Y \prec Y'$ implies $(X,Y) \prec (X',Y')$}
 \item {Scaling invariance: $\lambda >0$ and $X \prec Y$ implies $\lambda X \prec \lambda Y$}
 \item {Splitting recombination: $X \sim (\lambda X,(1-\lambda) X)$}
 \item {Stability: If $(X,\epsilon Z) \prec (Y, \epsilon Z')$ then $X \prec Y$ for $\epsilon \rightarrow 0^{+}$. This means that a 'small' additional system $\epsilon Z$ cannot perturb ordering of two systems $X, Y$.}
\end{itemize}
Up to now the relation $\prec$ is a partial order, however it can be made a total ordering by the following Comparison 'Hypothesis' that can be proved using the definition of a simple systems and the Zeroth Law of Thermodynamics  \cite{EntropyOrdering2}
\begin{Definition}
We say that the \emph{Comparison Hypothesis} (CH) holds for a state-space $\Gamma$ if all pairs of states in $\Gamma$ are comparable.
\end{Definition}

These assumptions/hypothesis imply the existence of entropy
\begin{Theorem} \cite{EntropyOrdering2, EntropyOrdering}
 A function $S:\Gamma \rightarrow \mathcal{R}$ called entropy exists under assumption of the above axioms and Comparison Hypothesis, and fulfills:
\begin{itemize}
 \item {Monotonicity: $X \prec Y \quad \Leftrightarrow \quad S(X) \leq S(Y)$}
 \item {Additivity: $S(X,Y) = S(X) + S(Y)$}
 \item {Extensibility: $S(\lambda X) = \lambda S(X)$}
\end{itemize}
\end{Theorem}

The last issue is to make consistent all local entropies for subsystems and check if the global entropy can be defined. This is done in the following
\begin{Theorem}\cite{EntropyOrdering2, EntropyOrdering}
 Assume that CH holds for all compound systems. For each system $\Gamma$ let $S_{\Gamma}$ be some definite entropy function on $\Gamma$. Then there are constants $a_{\Gamma}$ and $B(\Gamma)$ such that the function S, defined for all states of all systems by affine transformation
 \begin{equation}
  S(X)=a_{\Gamma}S_{\Gamma}(X)+B(\Gamma),
 \end{equation}
 for $X \in \Gamma$ , satisfies additivity (2), extensivity (3), and monotonicity (1) in the sense that whenever $X$ and $Y$ are in the same state-space, then
 \begin{equation}
  X \prec Y \quad \Leftrightarrow \quad S(X) \leq S(Y).
 \end{equation}
\end{Theorem}

The total ordering $\prec$ of adiabatic processes and the existence of entropy $S$ that fulfills monotonicity (for simple systems) establishes a link with Theorem \ref{Th.SmoothMonotinicity} of smooth case. It is also a starting point to define entropy in terms of category theory, which will be the subject of the next section.

\section{Categorification}
In this section, we review some concepts from \cite{KyciaCategoryEntropy}. For background from category theory see \cite{CategoryGentleIntroduction} or \cite{CategoryTheoryForWorkingMathematicans}.

We will consider only a simple (i.e., not compound) systems for simplicity. This approach is based on the definition of Poset (pre-ordered set) as a category:
\begin{Definition}
A poset (pre-ordered set) $(P,\prec)$ is a set $P$ with order relation $\prec$. The arrow $x\rightarrow y$ for $x,y \in P$  exists, by definition, when $x \prec y$. 
\end{Definition}
We will use the definition for $\prec$ being a total order since this is the case for entropy from previous sections. Then the ordering relation/the arrow $x\rightarrow y$ is defined only when $y$ is adiabatically accessible from $x$.

If scaling of the system is considered, then instead of Poset, the G-Poset category has to be considered \cite{GPoset}. The first step is to define a set with group action - a G-Set \cite{GSet} - that accommodates the space of states $\Gamma$ from the previous sections:
\begin{Definition}
System space is the object of the G-Set category, i.e., $\{\Gamma,(\mathbb{R}^{+},\cdot,1)\}$, where the multiplicative group acts on the set $\Gamma$.
\end{Definition}

In the next step, the definition of G-Poset can be adapted for $P=\Gamma$ - the space of states from the previous section - to define the system with entropy. Under the assumption from the previous section, on the poset the ordering is induced by the entropy $S:\Gamma \rightarrow \mathbb{R}$, and therefore we can define
\begin{Definition} \cite{KyciaCategoryEntropy}
The \emph{entropy system} is the object of G-Pos category, which objects are $\mathcal{G}=(\Gamma, \preccurlyeq)$, with preserving ordering group $(\mathbb{R}^{+},\cdot,1)$ action\footnote{If for $X,Y \in \Gamma$ there is $X\preccurlyeq Y$, then for $\lambda \in \mathbb{R}^{+}$ there is $\lambda X \preccurlyeq \lambda Y$.}, where the (partial or) total order is given by the entropy function $S:\Gamma \rightarrow \mathbb{R}$.
\end{Definition}

Hereafter we restrict ourselves only to Posets for simplicity. For the general case of G-Posets, see \cite{KyciaCategoryEntropy}.

Up to now, this is only rephrasing of the previous section in terms of 'abstract nonsense', and it does not introduce anything new. The situation, however, changes when we consider more than one entropy system. In this case, we have two or more posets that can represent different (and not necessary originating from thermodynamic) entropy systems. We can ask what is minimal mapping (Functors between these Posets) that preserves ordering, and therefore entropies that introduce these orderings. It occurs that the minimal 'relation' that preserves these orderings in both directions is the Galois connection \cite{CategoryGentleIntroduction, GaloisConnectionDefinition4}, which can be seen as a basic example/a 'prototype' of adjoint functors. The Galois connection rewritten in terms of orderings induced by entropy functions has the following form

\begin{Definition} \label{Def.LandauerConnection} \cite{KyciaCategoryEntropy} \textbf{The Landauer connection and Landauer's functor} \\
 Entropy system $\mathcal{G}_{1}=(\Gamma_{1}, S_{1} )$ is implemented/realized/simulated in the entropy system $\mathcal{G}_{2}= (\Gamma_{2}, S_{2})$ when there is a Galois connection between them, namely, there is a functor $F:\mathcal{G}_{1} \rightarrow \mathcal{G}_{2}$ and a functor $G:\mathcal{G}_{2} \rightarrow \mathcal{G}_{1}$ such that $F \dashv G$. 
 
 In terms of the entropy it is given as
 \begin{equation}
  S_{2}(Fc) \leq S_{2}(d) \Leftrightarrow S_{1}(c) \leq S_{1}(Gd).
  \label{Def.LandauersConnection}
 \end{equation}
We name the functors $F$ and $G$ the Landauer's functors.
\end{Definition}

The Galois connection usually appears in logical/model theory considerations when we have a Poset of some axioms, and we implement them on a Poset of models that realize these axioms \cite{CategoryGentleIntroduction}. The ordering is then provided by the 'strength' of axiom and model. In this vein, we can use the Landauer's connection to relate some abstract entropy model with its implementation on the physical system with thermodynamical entropy. If such a connection between these two levels model-realization exists, then the change in entropy at the level of the model is transferred through the Ladauer's connection to the change in entropy in the physical realization level. This was the original idea of Landauer \cite{Landauer, LandauerExplained}, who deduced that any irreversible logical operation at the level of Shanon-entropic system generates a physical heat. In terms of the Landauer's connection this heat is generated by the change of entropy in the physical part of the device that implements a logical system. Therefore, the categorical approach makes a sharp distinction, in which part of the compound entropic system such Landauer's heat is generated. This result also explained Maxwell's demon paradox \cite{KyciaCategoryEntropy}.

This sketch presents only one application of the connection. More details and examples from physics, computer science, and biology can be found in \cite{KyciaCategoryEntropy}.

\section{Summary}
In this paper, we presented the road from entropy in terms of thermodynamics to its categorification. We started from the foundations of thermodynamics and entropy that rely on contact structure. Having understood the motivation, the axiomatic approach to entropy was presented, which emphasizes the ordering of equilibrium states by adiabatic processes. Finally, this ordering was used to reformulate the system with entropy in terms of pre-ordered sets - Posets. Two such Posets can be Galois connected by functors that preserve orderings, and therefore entropies. This connection can be used in various interesting contexts.

\section*{Acknowledgments}
I would like to thank Valentin Lychagin for pointing me out this interesting subject and fruitful discussions. This overview was written thanks to the encouragement of Pasha Zusmanovich and warm, positive feedback of Lino Feliciano Res\'{e}ndis Ocampo. I am also grateful to Referees for their detailed and vital suggestions that help to improve the manuscript. This research was supported by the GACR grant GA19-06357S and Masaryk University grant MUNI/A/1186/2018. I also thank the PHAROS COST Action (CA16214) for partial support.

\appendix

\section{Differential forms}

The mathematical structure underlying equilibrium thermodynamics is the theory of differential forms on contact space and their integrability. This section outline the theory, and the interested reader is referred to various sources, including \cite{Edelen}. All theorems are local, which is convenient for applications. Therefore we restrict ourselves to open subsets of Euclidean space, which are diffeomorphic to open subsets of a manifold $N$, which will have a (local) coordinate chart $(x^{1},\ldots, x^{dim(N)})$.

\subsection{Frobenius theorem}
The basic problem in exterior calculus is to check complete integrability of an exterior system $\{\omega_{1},\ldots,\omega_{n}\}$, that is the existence of a submanifold given locally by $n$ relations $\Phi:=\{g_{i}(x)=c_{i},i=1\ldots n\}$, for constants $c_{i}$, on which the exterior system vanishes $\Phi^{*}\omega_{i}=0$ for $i=1\ldots n$. This is given by 
\begin{Theorem}\cite{Edelen}
The exterior system $\{\omega_{1},\ldots,\omega_{n}\}$ is completely integrable iff there exists a nonsingular matrix $A_{ij}$ of 0-forms that $\omega_{i}=\sum_{j}A_{ij}dg_{j}$.
\end{Theorem}

For a system given by a 1-form $Q$ complete integrability means that there exists an integrating factor (nonsingular 0-form) $T$ such that $Q=TdS$. This fact is useful in defining entropy.

This can be reformulated in terms of differential ideals. We say that the set $I$ is the differential ideal defined by the set of 1-forms $\{\omega_{1},\ldots,\omega_{n}\}$ if and only if for $\eta\in I$ we have $\eta = \sum_{i}A_{i}\omega_{i}$ for 0-forms $A_{i}$. In these terms the Frobenius theorem controls complete integrability of the differential ideal defined by the exterior differential system, namely,
\begin{Theorem} \cite{Edelen}
 The ideal $I$ is integrable iff $dI \subset I$.
\end{Theorem}
This means that the ideal $I$ is closed under the exterior derivative, i.e., $d\eta \in I$ if $\eta \in I$. This also means that $d\omega_{i}=\sum_{j} A_{ij}\omega_{j}$ for 0-forms $A_{ij}$, or $d\omega_{i}\wedge \omega_{1}\wedge\ldots\wedge\omega_{n}=0$ for $i=1\ldots n$.

An alternative version of the Frobenius theorem is formulated for distributions. Define the vector space $D=Span(Ker(\omega_{1}),\ldots,Ker(\omega_{n}))$. This means that at each point of the space we define a vector subspace, and we are asking if these subspaces are tangent to some submanifold that is an integral manifold of the distribution $D$. Then the Frobenius theorem has the form
\begin{Theorem}\cite{Edelen}
 The distribution $D$ is integrable iff $[D,D]\subset D$.
\end{Theorem}
This means that taking all possible vectors from the distribution (which can be associated with infinitesimal transformations on the manifold), by making their brackets, we cannot get new vectors (infinitesimal transformations) that are outside the distribution $D$. This observation gives the Caratheodory's theorem on accessibility:
\begin{Theorem} \cite{GeometryOfPhysicsFrankel, Edelen} 
If in the neighborhood of any point there are points not accessible by paths which have tangent vectors in the distribution, then the 1-form $\theta$ is integrable ($\theta \wedge d\theta=0$). 
\end{Theorem}

Summing up, if the distribution/exterior differential ideal is integrable, then it defines a foliation of the manifold/holonomic constraint. However, this statement is local. Foliation can behaves 'pathologically' forming, e.g.,  initial submanifold \cite{NaturalOperationsInDifferentialGeometry}. For defining the global structure of the leaves, and to assure that they are proper submanifolds, some additional work must be done.

The Frobenius theorem is useful in proving the existence of entropy, which is the Second Law of Thermodynamics.

\subsection{Darboux theorem}
The next important theorem is the Darboux theorem that describes local canonical form of the differential 1-form defining contact and symplectic structures on manifold. We present only version for contact form
\begin{Theorem}\cite{Edelen}\\
For a 1-form $\omega$ fulfilling $\omega\wedge(d\omega)^{n}\neq 0$ and $(d\omega)^{n+1}=0$ there exists $n+1$ local functions $\{X_{i}(x)\}_{i=0}^{n}$ and $n$ functions $\{P_{i}(x)\}_{i=1}^{n}$ on the manifold $M$ with coordinates $x^{i}$ such that the form $\omega$ has representation
\begin{equation}
 \omega=dX_{0}+\sum_{i=1}^{n}P_{i}dX_{i}.
\end{equation}
\end{Theorem}
These functions can be used to introduce new coordinates on the manifold in which $\omega$ has a simpler form.

\subsection{Contact structure}

We define
\begin{Definition}
 The pair $(M,\theta)$ where $M$ is odd dimensional manifold of dimension $2n+1$ and $\theta$ is non-degenerate 1-form that fulfills $\theta\wedge (d\theta)^{n}\neq 0$, is called contact manifold.
\end{Definition}
Since $dim(M)=2n+1$ and $deg(\theta\wedge (d\theta)^{n})=2n+1$ therefore $(d\theta)^{n+1}=0$. We can use the Darboux theorem to conclude that locally we can introduce coordinates that $\theta=dX_{0}+\sum_{i=1}^{n}P_{i}dX_{i}$.

The contact space and the contact form are in thermodynamics introduced by the First Law of Thermodynamics.

The contact structure is solvable by a submanifold $\Phi$ that fulfills $\Phi^{*}\theta=0$. We can ask about the maximal dimension of such submanifold. This is controlled by the following:
\begin{Theorem}\cite{LychaginContactGeometry}
Every maximal submanifold in a $2n+1$ dimensional contact manifold $M$ has dimension $n$ and is called Legendre submanifold.
\end{Theorem}

\subsection{Contact structure vs Jet space}
We finish this overview of differential geometry by discussing the rudiments of jet spaces. This presentation is mainly based on \cite{KushnerSlovakNonlinear, LychaginContactGeometry}.

Consider a $n$ dimensional manifold $N$ and smooth functions on the manifold $C^{\infty}(N)$. In local coordinates $(x_{1},\ldots x_{n})$ define the ideal 
\begin{equation}
 \mu_{a}^{k}:=\left\{ f \in C^{\infty}(N) \quad | \quad \frac{\partial ^{|\sigma|}f}{\partial x^{\sigma}}(a)=0, 0<|\sigma|<k \right\},
\end{equation}
where multiindices $\sigma=(\sigma_{1},\ldots,\sigma_{n})$, $|\sigma|=\sum_{i=1}^{n}\sigma_{n}$, and $\frac{\partial ^{|\sigma|}f}{\partial x^{\sigma}}:=\frac{\partial ^{|\sigma|}f}{\partial (x^{1})^{\sigma_{1}}\ldots (x^{n})^{\sigma_{n}}}$.

Now define the k-th jet of functions at $x=a$ as the quotient
\begin{equation}
 J^{k}_{a}(N):={C^{\infty}(N)}\diagup{\mu_{a}^{k+1}}.
\end{equation}
The equivalence classes $[f]^{k}_{a}\in J^{k}_{a}(N)$ represent the functions that have the same derivatives/contact at $x=a$ up to order $k$, in other words, their Taylor series at $x=a$ agree up to order $(x-a)^{k}$. For example for $dim(N)=1$, $[x]^{i}_{0}=[\sin(x)]^{i}_{0}$ for $i=0,1,2$ but disagree for $i=3$.

The k-jet of functions on $N$ is defined as
\begin{equation}
 J^{k}(N)=\bigcup_{a\in N}J^{k}_{a}(N).
\end{equation}
It is a fiber bundle $\pi: J^{k}(N)\rightarrow N$ with the obvious projection.

We can now describe  $J^{k}(N)$ locally by defining the ideal of 1-forms (the Cartan distribution). For simplicity consider $J^{1}(N)$. The local coordinates are $(x^{i},y,y_{i})$ where the new coordinates $p_{i}$ are associates with derivatives $\frac{\partial}{\partial x^{i}}$ of functions $C^{\infty}(N)$ by the Cartan distribution
\begin{equation}
 \omega=dy-p_{i}dx^{i}.
\end{equation}
The distribution is nonintegrable since $\omega\wedge d\omega \neq 0$. In addition, $\omega\wedge (d\omega)^{n}\neq 0$ and $dim(J^{1}(N)=2n+1$. This is exactly the local form from the Darboux theorem and also from the local definition of a contact form. Therefore the contact space $M$ of dimension $2n+1$ is exactly the 1-jet of smooth functions on $N$.

The sections of the jet bundle $s:N \rightarrow J^{1}(N)$ are called holonomic sections or 1-graphs (in case of $J^{k}(N)$ are called k-graphs) if 'differential' coordinates are derivatives, i.e.,
\begin{equation}
 x^{i}(s)=x^{i}, \quad y(s)=f, \quad p_{i}(s)=\frac{\partial f}{\partial x^{i}}.
\end{equation}
We can note that the section is a holonomic section iff it is a Legendre submanifold \cite{Lychagin_HolonomicAreLegendre}. This means that we can describe a Lagrange submanifold by a function $y(s)=f(x^{1},\ldots, x^{n})$ and then all $p$ coefficients in the Cartan distribution or $P$ coefficients in a contact form are derivatives
\begin{equation}
 \left\{ 
 \begin{array}{l}
  p_{1}=p_{1}(x_{1},\ldots,x_{n})=\frac{\partial f}{\partial x_{1}} \\
  \ldots \\
  p_{n}=p_{n}(x_{1},\ldots,x_{n})=\frac{\partial f}{\partial x_{n}}.
 \end{array}
\right.
\end{equation}

Symmetries of contact structure are such transformations that preserve the Cartan distribution \cite{KushnerSlovakNonlinear, LychaginContactGeometry}. For a diffeomorphism $\phi:J^{1}(N)\rightarrow J^{1}(N)$ the following condition ensures that it is a contact symmetry:
\begin{equation}
 \phi^{*}\omega = \lambda_{\phi}\omega,
\end{equation}
where $\lambda_{\phi}$ is some smooth non-vanishing function on $J^{1}(N)$. This condition shows that the kernel of $\phi^{*}\omega$ is the same as the kernel of $\omega$ - they define the same contact distribution.

Apart of simple symmetries like translation $(x^{i},y,p_{i})\rightarrow (x^{i}+\alpha^{i},y+\beta,p_{i})$ the most important symmetry in thermodynamics is the Legendre transformation:
\begin{equation}
 (x^{i},y,p_{i})\rightarrow (p_{i}, y-x^{i}p_{i},-x^{i}),
\end{equation}
that interchange $x^{i}$ with corresponding $p_{i}$.

$J^{1}(N)$ is sufficient for thermodynamics, however for general theory of jet spaces consult \cite{LychaginContactGeometry, KushnerSlovakNonlinear, NaturalOperationsInDifferentialGeometry}.




\end{document}